\documentclass[onecolumn,aps,12pt]{revtex4}
\usepackage{amsmath,graphicx}

\newcommand{\tgf} {{2\gamma}}
\newcommand{\fgf} {{4\gamma}}
\newcommand{\ptgf}{{\pi^0 \rightarrow 2\gamma}}
\newcommand{\pfgf}{{\pi^0 \rightarrow 4\gamma}}

\newcommand{\pfg}{$\pfgf$~}

\newcommand{\bea}{\begin{eqnarray}}
\newcommand{\eea}{\end{eqnarray}}

\begin{document}

\pagestyle{headings}


\title{Quark loop contribution to \pfg}

\author{Yu.M.~Bystritskiy}
\email{bystr@theor.jinr.ru}
\affiliation{Joint Institute for Nuclear Research, 141980 Dubna,
Russia}
\author{V.V.~Bytev}
\affiliation{Joint Institute for Nuclear Research, 141980 Dubna,
Russia}
\author{E.A.~Kuraev}
\email{kuraev@theor.jinr.ru}
\affiliation{Joint Institute for Nuclear Research, 141980 Dubna,
Russia}

\date{\today}

\begin{abstract}

    We find the contribution of constituent quark loop mechanism to the branching ratio
    of \pfg process
    to be $B_\fgf^{hadr} = \Gamma_\pfgf/\Gamma_\ptgf \approx 5.45 \cdot 10^{-16}$ for the
    reasonable choice of constituent quark mass $m \approx 280~\mbox{MeV}$. This result is in
    agreement with vector-dominance approach result obtained years ago. Thus the main
    contribution arises from QED mechanism $\pi^0 \rightarrow \gamma(\gamma^*)
    \rightarrow \gamma (3\gamma)$ including light-light scattering block with electron
    loop. This contribution was investigated in paper of one of us and gave
    $B_\fgf^{QED} \sim 2.6 \cdot 10^{-11}$.

\end{abstract}

\maketitle


\section{Introduction}

In the program of investigation of CP-violating effects one of directions is the search of
the "forbidden" process \cite{McDonough:1988nf,Baranov}:
\bea
    \pi^0 \rightarrow 3 \gamma.
\eea
An important background to it is the allowed decay:
\bea
    \pi^0 \rightarrow 4 \gamma.
\eea
A realistic estimation of the differential width of it is needed. Two types of contributions --
hadronic and electromagnetic -- as a relevant mechanisms of \pfg must be considered.

It was suggested in \cite{Schult:1992te} (1972) and confirmed in \cite{Bratkovskaya:1995hk}
(1995) that the main contribution have an electromagnetic nature:
\bea
    B^{QED}_{4\gamma} =
    \frac{\Gamma_\pfgf}{\Gamma_\ptgf}
    \sim 2.6 \cdot 10^{-11}.
\eea
As for hadronic part - two quite different predictions was done: paper \cite{Parashar:1975mm}
gave $B^{hadr}_{4\gamma} \sim 10^{-9}$ and paper \cite{Schult:1992te} result is
$B^{hadr}_{4\gamma} \sim 10^{-14}$.
Significantly lower bound for branching $B^{hadr}_{4\gamma} \sim 7.1\cdot 10^{-18}$
was quite recently obtained in frames of chiral perturbation theory in
\cite{Liao:1998gk}.

This contradiction is the motivation of our investigation.

It's known the duality property in description of strong-interaction phenomena:
\begin{itemize}
    \item The meson and barion (hadrons) approach;
    \item Quark-gluon approach.
\end{itemize}
In this paper we estimate (using the quark-gluon approach) the \pfg width in the
model with constituent quark loop.

\section{APPROACH BASED ON PCAC AND VECTOR-DOMINANCE HYPOTHESES}

Considering the process \pfg within the Vector-Dominance model (VDM) leads to the
following Feynman diagram (FD) Fig.~\ref{Fig1}. The amplitude which corresponds to this FD is
\cite{Schult:1992te}:
\begin{figure}
\includegraphics[scale=1]{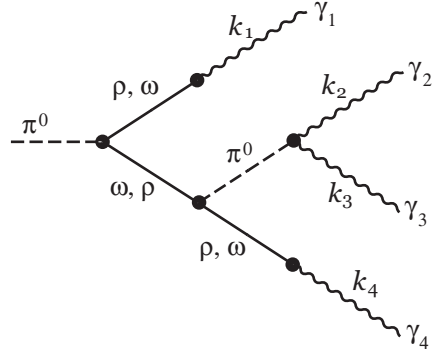}
\caption{Diagrams of hadronic contribution within Vector-Dominance model (VDM).}
\label{Fig1}
\end{figure}
\bea
    M_\fgf &=& \frac{M_{2\gamma}}{M}
          \left[
              \frac{g_{\rho\pi\gamma}^2}
              {(k_2+k_3+k_4)^2-m_\rho^2}
              +
              \frac{g_{\omega\pi\gamma}^2}
              {(k_2+k_3+k_4)^2-m_\omega^2}
          \right]
          \frac{1}{(k_3+k_4)^2-M^2}
          \times\nonumber \\
    &\times& \varepsilon_{\mu\nu\sigma\lambda}~
    e_3^\mu k_3^\nu e_4^\sigma k_4^\lambda~
    \cdot
    \varepsilon_{\alpha\beta\gamma\delta}~
    (k_2+k_3+k_4)^\beta e_1^\gamma k_1^\delta
    \cdot
    \varepsilon^\alpha_{\beta'\gamma'\delta'}~
    (k_2+k_3+k_4)^{\beta'} e_2^{\gamma'} k_2^{\delta'} +
    \nonumber\\
    &+& \mbox{11~terms obtained by permutations},
    \label{VDM}
\eea
\noindent where $\{e_i, k_i\}$ -- is the polarization vector and
momentum of $i$-th photon; $M$ -- is the mass of pion and
$M_{2\gamma}$ -- is the amplitude of
$\pi^0\rightarrow2\gamma$ decay.
The coupling constants are fixed by VDM to be:
\bea
    g_{v \pi\gamma} = \frac{g_{\pi\omega\rho} \lambda_v}{M^3}
\eea
\noindent where $\lambda_\rho =
g_\rho$,\quad$\lambda_\omega=g_\omega/\sqrt{2}$,\quad$g_\rho^2=2F_\pi^2
m_\rho^2$,\quad$g_{\pi\omega\rho}^2/4\pi=0.51$~(see \cite{Gell-Mann:1962jt}),\quad
$F_\pi = 94~\mbox{MeV}$~ and:
\bea
    g_\omega^2=\left\{
                \begin{array}{rcl}
                &&0.43~g_\rho^2, \qquad
                \mbox{Das, Mathur, Okubo \cite{Das:1967ek}.}
                \\
                &&0.23~g_\rho^2, \qquad
                \mbox{Oakes, Sakurai \cite{Oakes:1967}.}\\
                \end{array}
                \right.
\eea
\noindent Estimating the branching ratio contribution which comes
from amplitude (\ref{VDM}) one gets \cite{Schult:1992te}:
\bea
    B^{hadr}_\fgf\ \le
                \left\{
                \begin{array}{rcl}
                &&8.6 \cdot 10^{-16}, \qquad
                \mbox{Das, Mathur, Okubo.}
                \\
                &&7.0 \cdot 10^{-16}, \qquad
                \mbox{Oakes, Sakurai.}\\
                \end{array}
                \right.
\eea

\noindent In paper \cite{Parashar:1975mm} the photons identity was not
taken into account (thus destructive interferences were lost).
And this gave the strongly overestimated result $B^{hadr}_\fgf \sim
10^{-9}$.

\section{Approach of constituent (heavy) quark loop}

In this paper we consider the mechanism with the constituent quark
loop. First we need to determine the pion-quark coupling constant $g$.
To do this we write out the $\pi^0 \to 2\gamma$ amplitude
within quark-loop approach assuming the pion wave function to be
\bea
    \pi^0 = \frac{1}{\sqrt{2}}(u {\bar u} - d {\bar d}).
\eea
\noindent This gives (see Fig.~\ref{Fig2} a):
\begin{figure}
\includegraphics[scale=1]{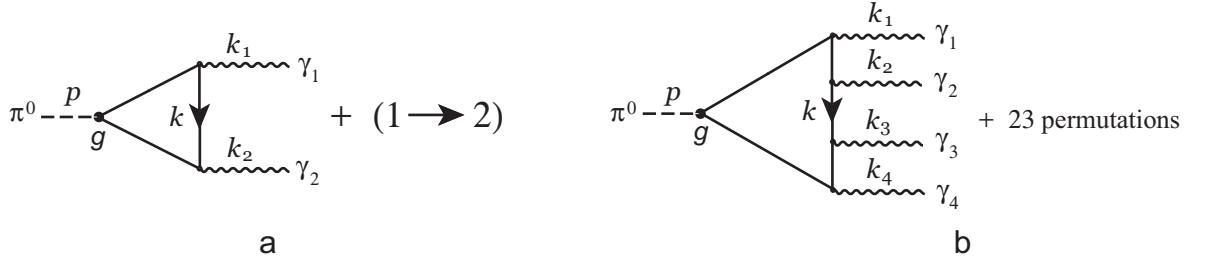}
\caption{Diagrams of hadronic contribution within constituent
quark loop approach.} \label{Fig2}
\end{figure}
\bea
    M_\tgf = 2 i g \frac{\alpha m}{\pi M^2}~
                    \frac{N_2}{\sqrt{2}}~
                    F(z) (e_1 e_2 k_1 k_2),
                    \label{QuarkLoopAmplitudeForPiTo2Gamma}
\eea
\noindent where $(e_1 e_2 k_1 k_2) \equiv
\varepsilon_{\mu \nu \alpha \beta} e_1^\mu e_2^\nu k_1^\alpha k_2^\beta$,~
$\alpha = 1/137$ - is fine-structure constant, the color-charge factor is
$N_2 = 3 \cdot \left( \left(\frac{2}{3}\right)^2 -
\left(\frac{1}{3}\right)^2 \right) = 1$,
$m, M$ -- constituent quark mass and neutral pion mass correspondingly,
$z = M^2/m^2$ and
$F(z) = \int\limits^1_0 \frac{dx}{x}\ln(1-z x(1-x))$. We use constituent
quark mass equal to $m=280~\mbox{MeV}$ (according to the analysis performed in
paper \cite{Volkov:1997dx}), thus $z \approx 0.23$ and $F(z) \approx -0.118$.

Comparing (\ref{QuarkLoopAmplitudeForPiTo2Gamma}) with current-algebra
($CA$) result:
\begin{gather}
    M_\tgf^{CA} = \frac{\alpha}{2\pi F_\pi}(e_1 e_2 k_1 k_2),
    \qquad
    \Gamma_\tgf = \frac{\alpha^2}{2^6 \pi^3} \frac{M^3}{F_\pi^2}
                    \approx 7.4~\mbox{eV},
\end{gather}
\noindent with $F_\pi=94~\mbox{MeV}$, we obtain for pion-quark coupling
constant $g = 2.065$.

Now we start to estimate the hadronic contribution to \pfg~decay
within quark-loop approach which gives us the set of FDs presented at
Fig.~\ref{Fig2} b.  The amplitude for the first FD has the form:
\bea
    M_\fgf^{(1)} &=& - g \alpha^2 \frac{N_4}{\sqrt{2}} \cdot I,
    \qquad
               I =
               \int \frac{d^4k}{i \pi^2}
               \frac{\mbox{Sp}}{A_1~A_2~A_3~A_4~A_5},
\eea
where $A_1 = k^2-m^2$, $A_2 = (k+k_2)^2-m^2$, $A_3 = (k-k_3)^2-m^2$,
$A_4 = (k-k_3-k_4)^2-m^2$, $A_5 = (k+k_1+k_2)^2-m^2$; $k,m$ - are the
momentum and mass of quark in the loop ($m=280~\mbox{MeV}$).
$\mbox{Sp} \equiv \mbox{Sp}\left[\gamma_5 (\hat k - \hat k_3 - \hat k_4 + m)
              \hat e_4 (\hat k - \hat k_3 + m) \hat e_3
              (\hat k + m) \hat e_2 (\hat k + \hat k_2 + m)
              \hat e_1 (\hat k + \hat k_1 + \hat k_2 + m)
    \right]$.
The color-charge factor for this case is $N_4 =
3 \cdot \left( \left(\frac{2}{3}\right)^4 -
\left(\frac{1}{3}\right)^4 \right) = \frac{5}{9}$.
Next we join the denominators of loop fermions propagators $A_i$:
\bea
    \frac{1}{A_1 \cdots A_5} &=&
    4! \int\limits_0^1 \frac{d\tau}{\left( A_1 x_1 +  \dots + A_5 x_5\right)^5},
\eea
where $d\tau = dx_1\dots dx_5~\delta(x_1 + \dots + x_5 - 1)$, and get
$A_1 x_1 +  \dots + A_5 x_5 = (k+b)^2-d$, where
$b = (k_1+k_2) x_5 + k_2 x_2 - k_3 x_3 - (k_3+k_4) x_4$,\quad
$d=m^2-M^2 a^2$,\quad
$a^2 = x_4 x_5 + x_3 x_5 (t_{13}+t_{23}) + t_{23} x_2 x_3 +
x_2 x_4 (t_{23}+t_{24}) + t_{34} x_4 (x_1+x_2) + t_{12} x_5 (x_1+x_3)$,\quad
$t_{ij}\equiv 2(k_i k_j)/M^2$.
Performing a shift of loop momentum ($k \rightarrow \chi + b$) and
calculating the loop trace $\mbox{Sp}[\dots]$ we can cancel even powers of new
integration momenta $\chi$ with denominators $(\chi^2-d)$. After that we may
use a well-known formula of integration:
\bea
    \int \frac{d^4\chi}{i \pi^2} \frac{1}{\left(\chi^2 - d\right)^n} =
    \frac{(-1)^n}{(n-1)(n-2)}\frac{1}{d^{n-2}}.
\eea
To perform the integration over Feynman x's we will expand the obtained
expression for amplitude over $z = M^2/m^2 \approx 0.23$, i.e. use formula:
\bea
    d^{-n} = m^{-2n}\left(
                            1 + n a^2 z + \frac{1}{2!} n(n+1) (a^2 z)^2 +
                            \frac{1}{3!} n(n+1)(n+2) (a^2 z)^3 + \dots
                    \right).
\eea
This makes the integration over x's trivial:
\bea
    \int d\tau~x_1^{n_1} x_2^{n_2} x_3^{n_3} x_4^{n_4}
    =
    \frac{n_1!~n_2!~n_3!~n_4!}{(n_1 + n_2 + n_3 + n_4 + 4)!}.
\eea
Thus we obtain the amplitude for first FD:
\bea
    M_\fgf^{(1)} &=& - g \alpha^2 \frac{N_4}{\sqrt{2}} \cdot I,
    \qquad
    I = \frac{1}{m}
            \left(
                B_0 + z B_1 + z^2 B_2 + z^3 B_3 + ...
            \right)
    \label{AmplitudeAfterAllIntegrations},
\eea
where $B_i$ - some rather complicated but known functions of $e_i$ and
$t_{ij}$. The total amplitude $M_\fgf$ is the sum of 24
terms obtained from (\ref{AmplitudeAfterAllIntegrations}) by permutations
of final photon legs (i.e. pairs $\{e_i,k_i\}$):
\bea
    M_\fgf = \sum\limits_{perm} M_\fgf^{(1)}.
    \label{TotalAmplitudeAsSumOverPermutations}
\eea
Performing this summation we see that:
\bea
    \sum\limits_{perm} B_0 = \sum\limits_{perm} B_1 =
    \sum\limits_{perm} B_2 = 0, \qquad \sum\limits_{perm} B_3 \ne 0.
\eea
This can be obtained by applying Schouten identity:
\bea
    (p_1 p_2 p_3 p_4)Q_\mu =
    (\mu p_2 p_3 p_4) (Q p_1) + (p_1 \mu p_3 p_4) (Q p_2) +
    (p_1 p_2 \mu p_4) (Q p_3) + (p_1 p_2 p_3 \mu) (Q p_4),
\eea
to antisymmetric Levi-Civita tensors $(\alpha\beta\gamma\delta)$ in
functions $B_i$.
Here we should notice that total amplitude $M_\fgf$ fulfills the
requirements of gauge invariance and Bose symmetry:
\bea
    k_1^\mu~M^\fgf_{\mu\nu\rho\sigma} =
    k_2^\nu~M^\fgf_{\mu\nu\rho\sigma} =
    k_3^\rho~M^\fgf_{\mu\nu\rho\sigma} =
    k_4^\sigma~M^\fgf_{\mu\nu\rho\sigma} = 0, \nonumber \\
    M^\fgf_{\mu\nu\rho\sigma}(k_1,k_2,k_3,k_4) =
    M^\fgf_{\nu\mu\rho\sigma}(k_2,k_1,k_3,k_4) = ...~.\nonumber
\eea
After squaring the amplitude $M_\fgf$ and summing over photons
polarizations $\{\lambda_1 \lambda_2 \lambda_3 \lambda_4\}$ we have:
\bea
    \sum\limits_{\lambda_1 \lambda_2 \lambda_3 \lambda_4}\left| M_\fgf \right|^2
    = \left( -g \alpha^2 \frac{N_4}{\sqrt{2}}\right)^2
    \frac{1}{225}~\frac{z^6}{m^2}~P,
    \qquad \mbox{with} \qquad
    P = \sum\limits_{perm} S_2,
\eea
where:
\bea
    S_2 &=&
\frac{1}{2} t_{12} \left(8 t_{13} \left(t_{14}^2 \left(t_{12} t_{13} +
                    t_{23} \left(-2 t_{14} + 3 t_{23}\right)\right) +
                \right.\right.\nonumber \\
                &&+\left.
                    \left(3 t_{12} t_{13} +
                    4 t_{23} \left(-2 t_{14} + t_{23}\right)\right) t_{24}^2 +
              6 t_{13} t_{24}^3\right) + 4 t_{13} t_{14} \left(t_{14} - t_{23}\right) t_{24} t_{34} -
        \nonumber \\
        &&-\left.32 t_{13} t_{24}^2 t_{34}^2 +
         \left(3 t_{12}^2 - 40 t_{13} t_{24}\right) t_{34}^3 +
        13 t_{12} t_{34}^4\right).
\eea
The \pfg~decay width $d\Gamma^{hadr}_\fgf$ has a form:
\bea
    d\Gamma^{hadr}_\fgf = \frac{1}{(2\pi)^8 2M} \int
    \sum\limits_{\lambda_1 \lambda_2 \lambda_3 \lambda_4}\left| M_\fgf \right|^2~
    d\phi_4,
\eea
where the phase volume $d\phi_4$ is considered in Appendix~
\ref{AppendixPhaseVolume}. To evaluate the hadronic contribution
to total decay width $\Gamma^{hadr}_\fgf$ we perform the numerical
phase-space integration and finally obtain:
\bea
    \Gamma^{hadr}_\fgf =
    \frac{M}{2^{11} \pi^6}~g^2 \alpha^4 \frac{N_4^2}{225}~
    z^7 {\cal J},
    \qquad \mbox{where} \qquad
    {\cal J} = \int~d{\tilde \phi_4}~P\approx 0.097
    \label{HadronicDecayWidth},
\eea
where the dimensionless phase volume $d{\tilde \phi_4}$ is defined
by relation $d\phi_4=\frac{\pi^2}{2}M^4 d{\tilde \phi_4}$. Thus our result
for hadronic contribution to total decay width is
$\Gamma^{hadr}_\fgf \approx 4 \cdot 10^{-15}~\mbox{eV}$.

\section{Conclusion}

Using the result listed above (see (\ref{HadronicDecayWidth}))
we estimate the hadronic contribution to branching of
\pfg decay as:
\bea
    B^{hadr}_\fgf = \frac{\Gamma^{hadr}_\fgf}{\Gamma_\tgf}
    \approx 5.45 \cdot 10^{-16}.
\eea
This value has the same order of magnitude as one
obtained in \cite{Schult:1992te}. It also is in a good agreement with
the recent result \cite{Silagadze:2003xm} obtained within the meson and
barion approach. But it contradicts the result of paper
\cite{Liao:1998gk} obtained by using chiral perturbation theory.

The result obtained in \cite{Tarasov} $B^{hadr}_\fgf \le 10^{-9}$ is in
strong contradiction to our estimate and QED result
\cite{Bratkovskaya:1995hk}. As for result of paper \cite{Parashar:1975mm} it
presumably needs a serious revision as well as amplitudes interferences
was not taken into account.

We should notice that our consideration were performed within QED.
As for QCD corrections they may be parameterized in the
variation of constituent quark mass $m$. For instance if
uncertainty of $m$ is about 5 MeV it can result in
$\delta B_{4\gamma}/B_{4\gamma} \sim 30 \%$. So these corrections can
be essential.

Finally we may conclude that the main contribution to \pfg~decay has
the QED nature, i.e. mechanism with light-light scattering block
with electron loop: $\pi^0 \to \gamma (\gamma^*) \to \gamma
(3\gamma)$. This contribution was investigated in paper of one of us
\cite{Bratkovskaya:1995hk} and appeared to be equal to
\begin{gather}
    B^{QED}_\fgf \approx 2.6 \cdot 10^{-11}.
\end{gather}

\appendix

\section{Phase volume}
\label{AppendixPhaseVolume}

\noindent The phase volume $d\phi_4$ for decay \pfg is:
\bea
        d\phi_4 =
        \frac{d^3k_1}{2\omega_1}\frac{d^3k_2}{2\omega_2}
        \frac{d^3k_3}{2\omega_3}\frac{d^3k_4}{2\omega_4}
              \delta^4\left( p - k_1 - k_2 - k_3 - k_4 \right),
    \quad p^2 = M^2,\quad k_i^2 = 0.
\eea
Being expressed in terms of energy fractions of photons
$y_i=\omega_i/M$ (which satisfy the identity
$y_1+y_2+y_3+y_4=1$) and cosines of photons momenta
orientations $C_{ij}=\cos(\widehat{\vec k_i \vec k_j})$ this phase
volume reads to be:
\bea
    \int d\phi_4~\dots &=&
              \frac{\pi^2}{2}~M^4
              \int\limits_0^{1/2} d y_1~d y_2~
              \int\limits_{0}^{2\pi} d \phi_3
              \int\limits_{-1}^1 d C_{12}~d C_{13}
              \frac{y_1~y_2~y_3}{A}~
              \dots,
\eea
where $A=1-y_1(1-C_{13})-y_2(1-C_{23})$.
The region of integration is determined by conditions:
\bea
&&0 < y_{1,2} < \frac{1}{2}, \qquad -1 < C_{ij} < 1, \nonumber \\
&&0 < y_3 = \frac{1}{2A}(1 - 2(y_1+y_2)+2y_1y_2(1-C_{12})) < \frac{1}{2}, \\
&&0 < y_4 = \sqrt{y_1^2+y_2^2+y_3^2+2y_1y_2C_{12}+2y_1y_3C_{13}+2y_2y_3 C_{23}} < \frac{1}{2},
\nonumber
\eea
while $C_{23}=C_{12} C_{13} + S_{12} S_{13} \cos(\phi_3)$, and
$S_{ij}=\sin(\widehat{\vec k_i \vec k_j}).$
The kinematical variables $t_{ij}$ can be expressed as:
\bea
\begin{array}{ll}
    t_{12} = 2~y_1~y_2~(1-C_{12}),                      &
    t_{23} = 2~y_2~y_3~(1-C_{23}),                      \\
    t_{13} = 2~y_1~y_3~(1-C_{13}),                      &
    t_{24} = 2~y_2~(y_2+y_4+y_1 C_{12}+y_3 C_{23}),     \\
    t_{14} = 2~y_1~(y_1+y_4+y_2 C_{12}+y_3 C_{13}),     &
    t_{34} = 2~y_3~(y_3+y_4+y_1 C_{13}+y_2 C_{23}).
\end{array}
\eea


\end{document}